%% file: main.tex
\documentclass[10pt,conference]{IEEEtran}
\usepackage{graphicx} 
\usepackage[utf8]{inputenc}
\usepackage{amsmath,amsfonts,amsthm}
\usepackage{algorithm}
\usepackage{algorithmic}
\usepackage{xcolor}
\usepackage{xspace}
\usepackage{xurl}
\usepackage{enumitem}
\usepackage{pifont}
\usepackage{multirow}
\usepackage{multicol}
\usepackage{makecell}
\usepackage[hidelinks]{hyperref}
\usepackage{tcolorbox}
\usepackage{subcaption}
\usepackage{caption}
\usepackage{textcomp}
\usepackage{listings}
\usepackage{balance}
\usepackage{fancyhdr}

\IEEEoverridecommandlockouts

\makeatletter
\long\def\@makecaption#1#2{\vskip\abovecaptionskip
  \sbox\@tempboxa{#1.~#2}%
  \ifdim \wd\@tempboxa >\hsize
    #1.~#2\par
  \else
    \hbox to\hsize{\hfil\box\@tempboxa\hfil}%
  \fi
  \vskip\belowcaptionskip}
\makeatother

\newcommand{\learntofix}{\textsc{learn2fix}\xspace}
\newcommand{\afterofix}{\textsc{isonoise}\xspace}
\newcommand{\grammartofix}{\textsc{Grammar2Fix}\xspace}

\title{Isolating Noisy Labelled Test Cases in Human-in-the-Loop Oracle Learning}

\author{\IEEEauthorblockN{Charaka Geethal Kapugama} \IEEEauthorblockA{\textit{Department of Computer Science} \\ \textit{Faculty of Science} \\ \textit{University of Ruhuna} \\ Sri Lanka \\ charaka@dcs.ruh.ac.lk}}
\date{December 2024}

\begin{document}
 
\maketitle

\thispagestyle{fancy}
\fancyhf{} 
\fancyfoot[l]{\small 979-8-1234-5678-9/25/\$31.00~\copyright~2025 IEEE}

\renewcommand{\headrulewidth}{0pt}

\begin{abstract}
    Incorrectly labelled test cases can adversely affect the training process of \emph{human-in-the-loop oracle learning} techniques. This paper introduces \afterofix, a technique designed to identify such mislabelled test cases introduced during human-in-the-loop oracle learning. This technique can be applied to programs taking numeric inputs. Given a compromised automatic test oracle and its training test suite, \afterofix first isolates the test cases suspected of being mislabelled. This task is performed based on the level of disagreement of a test case with respect to the others. An intermediate automatic test oracle is trained based on the slightly disagreeing test cases. Based on the predictions of this intermediate oracle, the test cases suspected of being mislabelled are systematically presented for relabelling. When mislabelled test cases are found, the intermediate test oracle is updated. This process repeats until no mislabelled test case is found in relabelling. \afterofix was evaluated within the human-in-the-loop oracle learning method used in \learntofix. Experimental results demonstrate that \afterofix can identify mislabelled test cases introduced by the human in \learntofix with over 67\% accuracy, while requiring only a small number of relabelling queries. These findings highlight the potential of \afterofix to enhance the reliability of human-in-the-loop oracle learning.
\end{abstract}
\vspace{1em}

\input{sections/introduction}
\input{sections/related_work}
\input{sections/methodology}

\input{sections/experimental_setup}
\input{sections/experimental_results}
\input{sections/discussion}
\input{sections/conclusion}

\bibliographystyle{IEEEtran}
\bibliography{references}
\end{document}

%% file: sections/introduction.tex
\section{Introduction}

Ensuring fault-free software is a key concern in software development. However, this task has become challenging due to the growing complexity of software systems. Manually testing an industrial-level software system is no longer practical. Therefore, software developers are shifting towards automated methods for testing software. 

In automating the software testing process, developing \emph{automatic test oracles}~\cite{weyuker1982testing} is essential. In software testing, a \emph{test oracle} is any procedure that distinguishes between the correct and incorrect behaviours of the \emph{system under test} (SUT)~\cite{barr2014oracle}. The work of Briand~\cite{briand2008novel} emphasises that the automation of test oracles is probably one of the most difficult problems in software testing. Also, the survey paper of Barr et al.~\cite{barr2014oracle} highlights that the problem of test oracle automation has been under-explored. The lack of techniques for test oracle automation poses a significant obstacle to implementing automated software testing. 

Human-in-the-loop automatic program repair focuses on fixing bugs in programs when the only available source of information about a test failure is the individual (the human) reporting the bug. Thus, these techniques can be applied to \emph{semantic bugs} or \emph{functional bugs}~\cite{tan2014bug}. \learntofix~\cite{geethal2023human}\cite{bohme2020human} and \grammartofix~\cite{grammar2fix} are examples of such techniques. In common, these techniques begin the process with an input exposing the bug (i.e., a failing input). Also, these techniques employ \textbf{human-in-the-loop oracle learning (HIOL)} methods~\cite{geethal2023human}\cite{grammar2fix}. Essentially, HIOL methods use fewer data points (20 - 50) to train automatic test oracles, as labelling a large number of data points by a human is impractical.

HIOL methods assume that the human correctly answers the labelling queries, since a person familiar with the expected behaviour of the program under test is involved in the labelling process. However, the human can make mistakes in labelling test cases as passing and failing. For numeric inputs, the journal extension of \learntofix~\cite{geethal2023human} highlights that incorrectly labelled test cases introduced during the learning process can adversely affect the quality of the generated automatic test oracles and patches. The simplest approach to resolve this issue is to verify the correctness of or relabel all the test cases in the training test suite. However, this is a significant overhead in terms of software testing.  

This paper introduces \afterofix, focusing on the aforementioned issues. The key objective of this approach is to \emph{identify incorreclty labelled test cases introduced in HIOL while using minimal relabelling queries}. 
This method can be applied to programs taking numeric inputs. In contrast to existing noise detection methods, \afterofix does not require a large data set to function properly. Also, it iteratively identifies mislabelled data points. However, \afterofix specifically focuses on data sets used for training automatic test oracles, in which test failures are rarely observed. 

Given a compromised automatic test oracle and its training test suite, \afterofix first isolates the test cases suspected of being mislabelled. For this task, \afterofix calculates the level of disagreement of each test case with respect to the others in the training test suite. To compute the disagreement, \afterofix employs a method based on the work of Bougelia et al.~\cite{bouguelia2018agreeing}. This method does not require an additional labelled test suite. To systematically present test cases for relabelling, \afterofix trains an intermediate automatic test oracle based on the slightly suspicious test cases in the training test suite and uses its predictions. When a mislabelled test case is found in relabelling, the intermediate test oracle is updated by retraining it with the correct label.  As the intermediate automatic test oracle is improved, more incorrectly labelled test cases can be revealed.  

To the best of available knowledge, \learntofix is the first and only human-in-the-loop automatic program repair technique for programs taking numeric inputs. Thus, its HIOL method was selected for the experiments conducted in this work. Several experiments were conducted using 552 program subjects from the Codeflaws\cite{Tancodeflaws} benchmark, which contains different types of real-world bugs. The experimental results demonstrate that:
\begin{enumerate}
    \item \afterofix can correctly identify the incorrectly labelled test cases introduced in the \learntofix active oracle learning. For the majority of subjects, \afterofix shows more than 67\% accuracy in identifying the incorrectly labelled test cases across the tested noisy label thresholds.
    \item \afterofix sends very few relabelling queries in the process of identifying incorrectly labelled test cases. This is a significant advantage over relabelling each test case in a training test suite individually.
    \item \afterofix significantly increases the likelihood of sending an incorrectly labelled test for relabelling.
\end{enumerate}

In summary, the \emph{main contributions} of this work are as follows.
\begin{enumerate}
    \item This work introduces a technique to identify incorrectly labelled test cases by the human in human-in-the-loop active oracle learning, without the need for an additional labelled test suite. 
    \item It presents an approach to reduce the search space for identifying the incorrectly labelled test cases introduced in human-in-the-loop active oracle learning. 
    \item A set of experiments is conducted on 552 buggy subjects to demonstrate that \afterofix is effective in identifying the incorrectly labelled tests introduced in the \learntofix active oracle learning.
\end{enumerate}

\textbf{Reproducibility.} To facilitate reproducibility, the implementation of \afterofix, the collected experimental data and the scripts have been made available at:

\textcolor{blue}{\url{https://github.com/charakageethal/ISONOISE}}
    

%% file: sections/related_work.tex
\section{Related Work}\label{sec:lit_review}

This paper was motivated by the work of Bougelia et al.~\cite{bouguelia2018agreeing} on active learning with noisy labels without crowdsourcing. This is a \emph{pool-based} active learning approach~\cite{settles2009active}, which assumes the existence of a set of unlabelled data points. In this work, Bougelia 's approach is extended to a \emph{stream - based} active learning approach, in which unlabelled data points (test cases) generated by mutational fuzzing~\cite{liang2018fuzzing}. The works of Zhang et al.~\cite{zhang2015bidirectional}, Northcutt et al.~\cite{northcutt2021confident} and Younesian et al.~\cite{younesian2021qactor} are also noisy (incorrect) label detection approaches for \emph{pool-based} active learning. The \emph{bi-directional active learning} (BDAL) approach presented by Zhang et al.~\cite{zhang2015bidirectional} provides an effective framework to deal with noisy labels encountered during active learning. The incremental oracle rectification process in \afterofix was inspired by this work. However, the BDAL method assumes the existence of a balanced data set at the beginning, as suggested by its experimental setup.  This assumption does not hold in the scenario where human-in-the-loop oracle learning is applied. The methods proposed by Northcutt et al.~\cite{northcutt2021confident} and Younesian et al.~\cite{younesian2021qactor} require large data sets (e.g. 1000K in~\cite{northcutt2021confident}) for better functioning. However, obtaining a large dataset is impractical in a human-in-the-loop environment.

The methods in~\cite{bouguelia2015stream} and~\cite{younesian2020active} can be applied to \emph{stream-based}~\cite{cacciarelli2024active} active learning approaches to detect noisy labels. Similar to~\cite{northcutt2021confident} and~\cite{younesian2021qactor}, these approaches also require large datasets for better performance, as implied by their experiments. Although the method in~\cite{younesian2020active} is a stream-based approach, the data are sent to the classifier in batches. 

Several approaches have been proposed for repairing/rectifying machine learning models. The works of Krishnan et al.~\cite{krishnan2017boostclean},  Cai et al.~\cite{cai2018b}, Li et al.~\cite{li2022hybridrepair} and Feng et al.~\cite{feng2020deepgini} are some example approaches. The methods proposed in~\cite{li2022hybridrepair} and \cite{feng2020deepgini} focus on \emph{deep learning}~\cite{guo2016deep} models and require additional datasets in the repair process. The work of Krishnan et al.~\cite{krishnan2017boostclean} considers \emph{domain value violations} in repairing a model. Therefore, it is necessary to have a strong background knowledge of the context in which the model is trained. The methods proposed by Cai et al.~\cite{cai2018b} and Barriga et al.~\cite{barriga2018automatic} can be considered domain-specific approaches.

%% file: sections/methodology.tex
\section{Methodology}

\begin{algorithm}
    \footnotesize
    \renewcommand{\algorithmicrequire}{\textbf{Input:}}
    \renewcommand{\algorithmicensure}{\textbf{Output:}}
    \newcommand{\algorithmicbreak}{\textbf{break}}
    \newcommand{\BREAK}{\STATE \algorithmicbreak}
    \caption{\footnotesize\textsc{isonoise}}
    \label{alg:afterofix}
    \begin{algorithmic}[1]
        \REQUIRE Automatic Test Oracle : $\mathcal{O}$
        \REQUIRE Test suite used to train $\mathcal{O}$ : $T$
        \STATE Let $D$ be the disagreement threshold.
        \STATE Let $f$ be the initial failing test ($f \in T$).
        \STATE $\textit{NoNoiseFound}=\texttt{False}$
        \STATE Exclude $f$ from $T$.~$T_{\textit{gen}} \gets T \setminus \{f\}$.
        \WHILE{$\textit{NoNoiseFound} \neq \texttt{True}$}
            \STATE $T_{S} \gets \emptyset$
            \FOR {$t \in T_{\textit{gen}}$}
                \IF{$\textsc{calculate\_disagreement}(t,\mathcal{O}) > D$}
                 \STATE $T_{S} \gets T_{S} \cup \{t\}$
                \ENDIF
            \ENDFOR
            \STATE Exclude $T_S$ from $T$,~ $T_{N} \gets T \setminus T_{S}$
            \STATE Sort $T_{S}$ in descending order based on the disagreement scores.
            \STATE $\mathcal{O}_{N} \gets \textsc{train\_classifier}(T_{N})$
            \FOR{$t_{s} \in T_{S}$}
                \STATE Let $h_{\textit{old}}$ be the previous human label of $t_{n}$
                \IF{$\mathcal{O}_{N}(t_{s})=\textit{Failing} \vee \mathcal{O}_{N}(t_{s}) \neq h_{\textit{old}}$}
                    \STATE Relabel $t_{s}$. Let $h_{\textit{new}}$ be the new label of $t_{s}$.
                    \IF{$h_{\textit{new}}\neq h_{\textit{old}}$} 
                        \STATE Noisy label detected.
                        \STATE Change the label of to $h_{\textit{new}}$
                        \STATE Retrain: $\mathcal{O} \gets \textsc{train\_classifier}(T)$
                        \STATE $\textit{NoNoiseFound} \gets \texttt{False}$
                        \BREAK
                    \ELSE
                        \STATE $\textit{NoNoiseFound} \gets \texttt{True}$
                    \ENDIF
                \ENDIF
            \ENDFOR
        \ENDWHILE
    \end{algorithmic}
\end{algorithm}

\begin{algorithm}[htbp]
     \footnotesize
    \renewcommand{\algorithmicrequire}{\textbf{Input:}}
    \renewcommand{\algorithmicensure}{\textbf{Output:}}
    \caption{\footnotesize \textsc{calculate\_disagreement}}
    \label{alg:disagreement}
    \begin{algorithmic}[1]
        \REQUIRE Test Case : $t$, Human Label of $t$ : $\mathcal{H}(t)$ 
        \REQUIRE Automatic Oracle : $\mathcal{O}$, Training test suite of $\mathcal{O}$ : $T$
        \STATE Let $N$ be the fuzzing iterations. 
        \STATE Exclude $t$ from $T$, $T' \gets T \setminus \{t\}$ 
        \STATE $\textit{disagreement}=0$
        \FOR{$i \gets 1$ to $N$}
            \STATE $t' \gets \textsc{mutate\_fuzz}(t)$
            \STATE Predicted Label $\mathcal{L} \gets \mathcal{O}(t')$
            \STATE $\mathcal{O}_T \gets \textsc{train\_classifier}(T'\cup\{t'_{\mathcal{L}}\})$
            \IF{$\mathcal{H}(t) \neq \mathcal{O}_T(t)$}
                \STATE $\textit{disagreement}\gets\textit{disagreement}+1$
            \ENDIF
        \ENDFOR
    \RETURN \textit{disagreement}
    \end{algorithmic}
\end{algorithm}

\afterofix takes an automatic test oracle ($\mathcal{O}$) and the test suite ($T$) used to train it - which contains some incorrectly labelled test cases - as input. The automatic test oracle is a \emph{binary classifier} and has been trained using a human-in-the-loop oracle learning (HIOL) technique. Given a test input and its corresponding output of the program under test, this oracle can classify the test case as either passing or failing. Also, such an oracle expresses the condition undofer which the bug is observed~\cite{geethal2023human}\cite{bohme2020human}. A test case $t$ in $T$ is of the form $t=\langle \vec{i},o \rangle$, where $\vec{i}$ is a vector of input variable values, and $o$ is the output produced by the program under test for $\vec{i}$. Both $\vec{i}$ and $o$ contain numerical values, and $\vec{i}$ has a fixed length. 

Algorithm~\ref{alg:afterofix} describes \afterofix. First, \afterofix computes the disagreement scores of test cases generated in human-in-the-loop oracle learning ($T_{\textit{gen}}$). Here, the initial failing test ($f$), from which the oracle learning processing was started, is excluded (Algorithm~\ref{alg:afterofix}-Line 4). The reason is that HIOL begins with a single failing test that is considered correctly identified. \afterofix uses Algorithm~\ref{alg:disagreement} (\textsc{calculate\_disagreement}) to compute the disagreement scores of each test case in $T_{\textit{gen}}$. The test cases with disagreement scores greater than the disagreement threshold ($D$) are added to the set $T_{S}$ (Line 8 \& 9 - Algorithm~\ref{alg:afterofix}). This approach separates test cases suspected of being incorrectly labelled from the others. The test cases in $T_{S}$ are sorted in descending order by their disagreement scores. Next, another automatic oracle ($\mathcal{O}_{N}$) is trained using the test cases that are not in $T_{S}$ (Line 14 - \textsc{train\_classifier}-Algorithm~\ref{alg:afterofix}). 

In the next step, each test case $t_{s} \in T_{S}$ is presented to $\mathcal{O}_{N}$ to predict its label. The label of $t_{s}$ (i.e. passing or failing) is \emph{relabelled} if: 
\begin{enumerate}[label=\roman*]
    \item $t_{s}$ is predicted as \emph{failing} by $\mathcal{O}_N$~\textit{\textbf{OR}}
    \item The label predicted by $\mathcal{O}_{N}$ for $t_{s}$ is different from the label previously given by the human ($h_{\textit{old}}$) 
\end{enumerate}
Algorithm~\ref{alg:afterofix} - Line 17). 
If the new label of $t_s$ ($h_{\textit{new}}$) is different from the previous one ($h_{\textit{old}}$), it indicates that the human incorrectly labelled $t_{s}$ previously. This incorrectly labelled test case negatively affects the quality of the automatic oracle ($\mathcal{O}$). Hence, $\mathcal{O}$ is retrained using $T$ after changing the label of $t_s$ to $h_{\textit{new}}$. This process continues until no incorrectly labelled test case is found in $T_{S}$. \textsc{train\_classifier} in both Algorithm~\ref{alg:afterofix} and~\ref{alg:disagreement} uses the same classification algorithm applied in the human-in-the-loop oracle learning technique. 
\subsection{Calculating the Disagreement Score of a Test Case}\label{meth:disgreement}

The method of computing the disagreement (Algorithm~\ref{alg:disagreement}) of test cases is based on the noisy label detection approach without crowd-sourcing by  Bougelia et al.~\cite{bouguelia2018agreeing}.  The key idea of Bougelia's approach is that if many classifiers (machine learning models) agree on a common label for a particular data point that differs from its queried label, then the data point can be suspected of being incorrectly labelled. The given automatic test oracle ($\mathcal{O}$) is a binary classifier.  Following this approach, Algorithm~\ref{alg:disagreement} creates several classifiers by slightly modifying the training test suite of $\mathcal{O}$ (i.e.,$T$).

Bougelia's approach~\cite{bouguelia2018agreeing} first excludes the data point for which the disagreement score is to be calculated from the training dataset. Algorithm~\ref{alg:disagreement} also excludes $t$ from the training test suite (Line 2), resulting in a new test suite, $T'$. Bougelia's approach~\cite{bouguelia2018agreeing} requires some additional data points, which are obtained from a pool of unlabelled data points assumed to be given. In this approach, \emph{mutational fuzzing}~\cite{liang2018fuzzing}\cite{manes2019art} is employed to generate these additional unlabelled data points. Algorithm~\ref{alg:disagreement} applies mutational fuzzing on $t$ to generate a new test case $t'$.
(Line 5). 
As numeric inputs are considered in this work, \textsc{mutate\_fuzz} employs \emph{arithmetic mutations}~\cite{manes2019art}. According to \cite{bouguelia2018agreeing}, the newly generated test case $t'$ by mutational fuzzing is presented to the initial automatic test oracle $\mathcal{O}$ to predict its label (Line 6). By adding $t'$ with its predicted label $\mathcal{L}$ (passing or failing) to $T'$, a new test suite is formed. Based on this test suite ($T'\cup\{t'_{\mathcal{L}}\}$), a new automatic oracle (a binary classifier) $\mathcal{O}_T$ is trained (Line 7). Then, $t$, for which the disagreement score is to be calculated, is presented to $\mathcal{O}_T$ to predict $t$'s label (Line 8). If this predicted label ($\mathcal{O}_T(t)$) differs from the label assigned by the human ($\mathcal{H}(t)$) previously, \textit{disagreement} is incremented by $1$ (Line 8 \& 9). This process continues for $N$ iterations. Finally, the disagreement score of $t$ is returned. 

\subsection{Incremental Isolation of Noisy Labelled Test Cases}

The preliminary studies of this work suggest that Algorithm~\ref{alg:disagreement} alone is insufficient to isolate incorrectly labelled test cases. The reason is that when there are fewer data points, classification algorithms might develop patterns based on incorrectly labelled data points. HIOL is an environment where fewer training data points are used~\cite{geethal2023human}\cite{grammar2fix}. 

To address the aforementioned issue, \afterofix employs a \emph{divide-and-conquer} approach~\cite{frosyniotis2003divide}. First, it separates highly disagreeing training test cases from slightly disagreeing ones (Algorithm~\ref{alg:afterofix} Line 8 
\& 9). This separation is performed based on a disagreement threshold ($D$). Next, a new classifier $\mathcal{O}_{N}$ is trained based on the initial failing test ($f$) and the slightly disagreeing test cases. The objective of this step is to train an intermediate classifier, presumably without any incorrectly labelled test cases. This classifier ($\mathcal{O}_{N}$) is not highly accurate.

The works of \cite{geethal2023human} and \cite{bohme2020human} show that failing test cases of a bug are observed less frequently than passing test cases in programs taking numeric inputs. Nonetheless, the ability to accurately identify test failures is an expected feature of a test oracle. For this reason, the labels of test cases predicted as \emph{failing} by $\mathcal{O}_{N}$ are sent for relabelling. Also, test cases whose labels disagree with the predicted labels by $\mathcal{O}_{N}$ are also subjected to relabelling (Algorithm~\ref{alg:afterofix} - Line 17). By doing this, it becomes possible to verify the reason for the disagreement: whether $\mathcal{O}_{N}$ is insufficiently trained or the test case was previously incorrectly labelled.  As incorrectly labelled test cases are corrected, the initial automatic test oracle ($\mathcal{O}$) improves. This is why the entire process is restarted whenever an incorrectly labelled test case is found.

%% file: sections/experimental_setup.tex
\section{Experimental Setup}
To evaluate \afterofix, the human-in-the-loop oracle learning technique used in \learntofix~\cite{geethal2023human}\cite{bohme2020human} was selected. \learntofix was originally developed as an automatic program repair technique~\cite{gazzola2017automatic}. However, it incorporates an oracle learning algorithm that systematically interacts with the human reporting a bug. To the best of available knowledge, this is the first and only human-in-the-loop oracle learning technique that learns automatic test oracles for semantic bugs in programs taking numeric inputs. \learntofix learns automatic test oracles as \emph{decision trees}~\cite{charbuty2021classification}. Therefore, the same decision tree learning algorithm used in \learntofix was applied in the \textsc{train\_classifier} function in Algorithm~\ref{alg:afterofix} and~\ref{alg:disagreement}.

\subsection{Research Questions}
\begin{enumerate}[leftmargin=*,label=\textbf{RQ.\arabic*}]
    \item\label{rq1} How accurately does \afterofix identify incorrectly labelled test cases?
    \item\label{rq2} How many relabelling queries does \afterofix require for incorrect label identification? Does the probability of sending an incorrectly labelled test case for relabelling increase compared to a random choice of test cases?
\end{enumerate}

\subsection{Experimental Subjects}\label{expr:subs}

To evaluate \afterofix and answer the research questions from the previous section, the \emph{Codeflaws}~\cite{Tancodeflaws} benchmark was selected based on the following criteria. 

\begin{enumerate}
    \item There should be a sufficiently large set of algorithmically complex programs. 
    \item There should be a variety of real-world defects that cause \emph{functional bugs}, i.e., programs producing incorrect or unexpected output for certain inputs. 
    \item For each subject, there should be a \emph{golden version}, i.e., a program producing the expected, correct output for a given input. To obtain the correct label for a test case, the subject's (buggy program's) output is compared with the golden version's output for the given input. If both outputs are different, the label of the test case is considered \emph{failing}; otherwise, it is considered \emph{passing}. 
    \item For each subject, there should be a manually constructed \emph{test suite}. In this test suite, there should be at least one failing test case, i.e., a test input for which the buggy program and the golden program produce different outputs. Otherwise, neither \afterofix nor \learntofix can be applied. 
    \item For each subject, there should be numeric test inputs of constant length. For each such test input, the program should produce a numeric output. Otherwise, the classification algorithm used in \afterofix and \learntofix cannot be applied.
\end{enumerate}

\textbf{Codeflaws} contains 3,902 faulty programs that belong to 40 real-world defects. These programs have been written in the C programming language. A total of \textbf{552 subjects} from this benchmark satisfy the last three (3-5) criteria and were chosen for evaluating \afterofix. The selected subjects belong to 34 defect categories. For each subject, there is a manually constructed test suite. \textbf{IntroClass} and \textbf{ManyBugs} benchmarks~\cite{LeGoues15tse} were excluded, as they do not meet the aforementioned selection criteria. 

\subsection{Setup and Evaluation}\label{expr:setup_eval}
 
For each subject, \learntofix is run on a randomly selected failing test case under a \emph{noisy label threshold} by \learntofix. A \emph{noisy label threshold} is the percentage of incorrectly labelled test cases from the allocated human query budget to \learntofix. These incorrectly labelled test cases are introduced at random points during the oracle learning process. Also, the human oracle ($\mathcal{H}$) is simulated in the experiments using the buggy and golden versions of the program (See criteria 3 in Section~\ref{expr:subs}).
In a noisy label, the \emph{correct label is simply inverted}.   
 
Next, \afterofix is applied to the generated automatic oracle and its training test suite, both produced by \learntofix. 
It is assumed that the relabelling queries are correctly answered in the experiments. The number of incorrectly labelled test cases identified by \afterofix, along with the number of relabelling queries sent in the process, is tracked. The above process is repeated for the 5\%, 10\% and 20\% noisy label thresholds. The reduction in oracle quality under these noisy label thresholds has been tested in the \learntofix journal extension~\cite{geethal2023human}. 

The data related to the incorrect label identification and relabelling queries are used to answer~\ref{rq1} and \ref{rq2}, respectively. 

For the experiments of this work, the following values were fixed. 
\begin{itemize}
    \item \emph{Timeout}. For each subject, 10 minutes were allocated for \learntofix oracle learning and \afterofix. 
    \item \emph{Labelling Query Budget of} \learntofix.  The maximum number of labelling queries to the human was set at 20. Also, no limit was set to the queries sent by \afterofix.
    \item \emph{Disagreement Threshold}. The disagreement threshold was set at 15 (i.e. $D=15$ in Algorithm~\ref{alg:afterofix}).
    \item \emph{Fuzzing Iterations}. The fuzzing iterations for calculating disagreement scores was set to 20. (i.e. $N=20$ in Algorithm~\ref{alg:disagreement})
\end{itemize}

The \emph{Timeout}, \emph{Labelling Query Budget} and \emph{Fuzzing Iterations} used the same values as those in the \learntofix journal extension~\cite{geethal2023human}. The \emph{Disagreement Threshold} was decided based on some pilot experiments. 

Related to \ref{rq1}, the accuracy of identifying incorrectly labelled test cases was calculated. Moreover, the accuracies of identifying incorrectly labelled failing (i.e., failing test cases incorrectly labelled as passing) and passing (i.e., passing test cases incorrectly labelled as failing) test cases were calculated separately. Regarding~\ref{rq2}, the number of relabelling queries sent by \afterofix was calculated. In addition, the probability of an incorrectly labelled test case being included in the relabelling queries sent by \afterofix was computed. 

To mitigate the impact of randomness and to gain statistical power for the experimental results,  each experiment was repeated 30 times per subject. 

%% file: sections/experimental_results.tex
\section{Experimental Results}\label{sec:res}
\subsection{\ref{rq1}~Accuracy of Detecting Noisy Labelled Test Cases}\label{sec:res_rq1}

\begin{figure}[htbp]
    \centering
    \includegraphics[width=\linewidth]{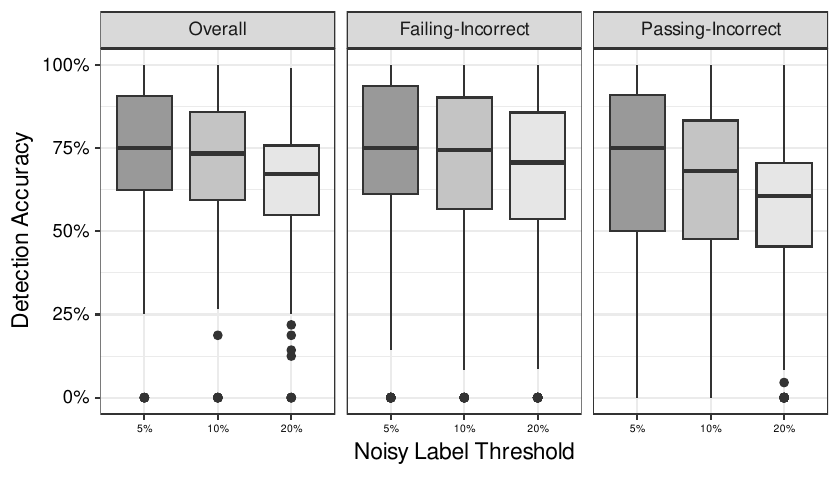}
    \caption{Detection Accuracy of \afterofix}
    \label{fig:noise_detection_accuracy}
\end{figure}

Fig.~\ref{fig:noise_detection_accuracy} shows the accuracy of \afterofix in detecting the incorrectly labelled test cases introduced during the \learntofix active oracle learning process. For each subject, the average values of \emph{Overall}, \emph{Failing-Incorrect} (i.e., failing test cases incorrectly labelled as passing) and \emph{Passing-Incorrect} (i.e., passing test cases incorrectly labelled as failing) were calculated over 30 runs. 

\begin{tcolorbox}
For the majority of subjects, \afterofix identifies the incorrectly labelled test cases with over 67\% accuracy under each noisy label threshold. Also, the median accuracy of identifying incorrectly labelled failing tests is over 70\%, and for incorrectly labelled passing tests, it is over 60\% under all the noisy label thresholds.
\end{tcolorbox}

According to these results, for most subjects, \afterofix can accurately identify the incorrectly labelled test cases introduced in the \learntofix human-in-the-loop active oracle learning. Also, the results suggest that \afterofix can more accurately identify incorrectly labelled failing tests (failing tests incorrectly labelled as passing). 

The other important fact indicated by the results is that the detection accuracy of incorrectly labelled tests decreases as the noise label threshold increases. As shown in Fig.~\ref{fig:noise_detection_accuracy}, the decrease in the median values of \emph{Failing-Incorrect} is smaller than that of \emph{Passing-Incorrect}.

\emph{Interpretation.} \afterofix uses only the automatic oracle and the test suite generated by \learntofix to detect incorrectly labelled test cases in the test suite. It does not require an additional labelled test suite to perform this task. Nevertheless, the results indicate that \afterofix is successful in identifying incorrectly labelled tests under various real-world bugs. 

The accuracy of identifying \emph{incorrectly labelled failing tests} is more than 70\% in most subjects under all noisy label thresholds. In \learntofix, the probability of labelling failing tests is maximized, as the failing inputs are essential to correctly learn the failure condition of a bug~\cite{geethal2023human}\cite{bohme2020human}. If a failing test is incorrectly labelled as passing, \learntofix might overlook some facets of the failure condition. Therefore, the ability of \afterofix to correctly identify incorrectly labelled failing tests is advantageous to \learntofix. 

As described before, the results show a drop in the detection accuracy as the noisy label threshold increases. The key reason is that as more incorrectly labelled tests are introduced, the quality of the automatic oracle given by \learntofix ($\mathcal{O}$) significantly decreases. Furthermore, when there are more incorrectly labelled tests, the classification algorithm might learn an alternative pattern that significantly differs from the target failure condition. As \learntofix uses a limited number of test cases (i.e., 21) in the training process, the drop in oracle quality is significant even with fewer incorrectly labelled test cases (This fact is highlighted in the \learntofix journal extension~\cite{geethal2023human}). A highly compromised oracle adversely affects the disagreement calculation process in Algorithm~\ref{alg:disagreement}, making it difficult to trace incorrectly labelled test cases. 

\textbf{Result.} \afterofix \emph{identifies incorrectly labelled test cases introduced in} \learntofix \emph{active oracle learning with high accuracy.} 

\subsection{\ref{rq2}~Relabelling Effort}\label{sec:res_rq2}

\begin{figure*}[htbp]
    \begin{subfigure}[t]{0.5\textwidth}
        \includegraphics[width=0.95\textwidth]{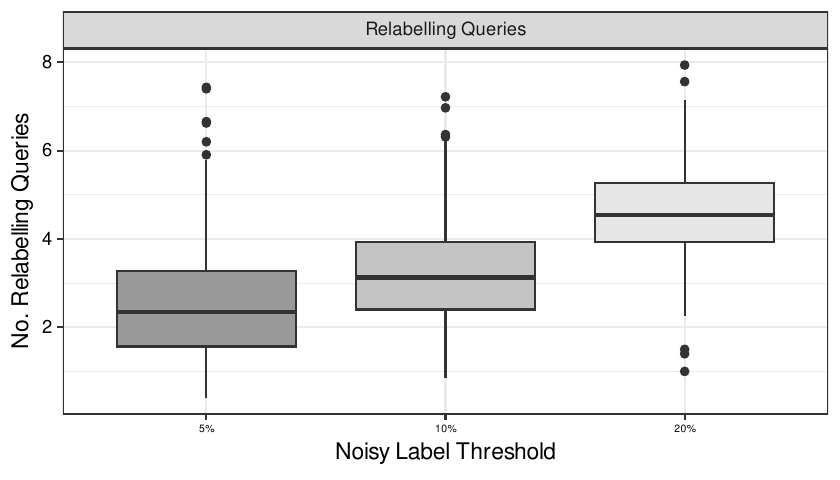}
        \caption{The number of relabelling queries}
        \label{fig:relabelling-effort-query}
    \end{subfigure}%
    ~
    \begin{subfigure}[t]{0.5\textwidth}
        \includegraphics[width=0.95\textwidth]{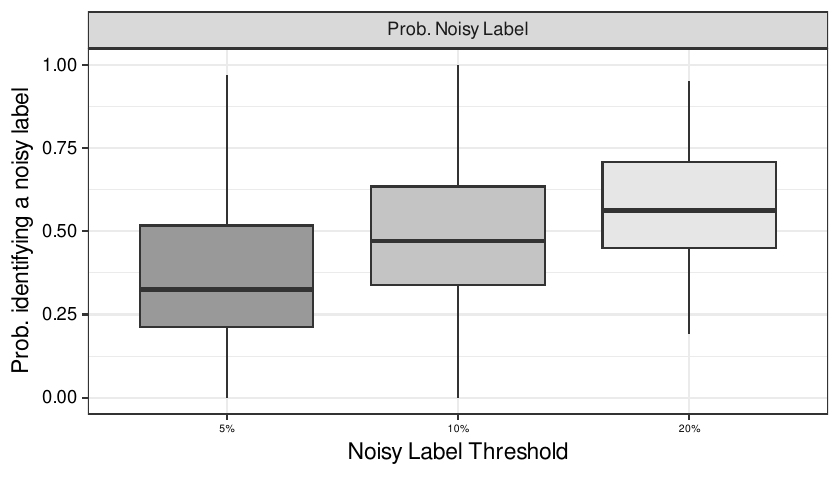}
        \caption{The probability of sending an incorrectly labelled test for relabelling}
        \label{fig:relabelling-effort-probability}
    \end{subfigure}%
    \caption{Relabelling Effort}
    \label{fig:relabelling-effort}
\end{figure*}

Fig.~\ref{fig:relabelling-effort-query} shows the number of relabelling queries sent under the three noisy label thresholds. Fig.~\ref{fig:relabelling-effort-probability} shows the probability of sending an incorrectly labelled test case for relabelling. For each subject, the average number of relabelling queries and mislabelled test cases was calculated over 30 runs to generate these graphs. 

\begin{tcolorbox}
    \afterofix sends only a few test cases for relabelling. Nevertheless, there is a higher probability of an incorrectly labelled test case being included in a relabelling query.
\end{tcolorbox}

According to the box-plots in Fig.~\ref{fig:relabelling-effort-query}, the median number of relabelling queries is slightly higher than the number of incorrectly labelled tests. Also, the median number of relabelling queries increases as the noisy label threshold increases. A similar pattern can be observed in the median probability of sending an incorrectly labelled test case for relabelling (Fig.~\ref{fig:relabelling-effort-probability}). In the 30\% noisy label threshold, this probability exceeds 0.5.

\emph{Interpretation.} The results suggest that \afterofix uses a small number of relabelling queries to identify incorrectly labelled test cases and to rectify automatic test oracles in most subjects. The lower relabelling effort suggests that \afterofix effectively reduces the search space needed to examine in relabelling.

When the entire allocated query budget is used by \learntofix, the probability of randomly picking an incorrectly labelled test is 0.05 at the noisy label threshold 5\%; 0.2 at 10\%; and 0.3 at 30\%. With \afterofix, this probability is significantly increased (Fig.~\ref{fig:relabelling-effort-probability}). Thus, \afterofix more frequently selects mislabelled test cases for relabelling.  The higher accuracy in detecting incorrectly labelled tests (Section~\ref{sec:res_rq1}) also supports this conclusion. 

\textbf{Result.} \afterofix \emph{effectively uses relabelling queries in exploring incorrectly labelled tests. Furthermore, incorrectly labelled tests are more frequently sent for relabelling}.

%% file: sections/discussion.tex
\section{Discussion and Future Work}
In \emph{human-in-the-loop oracle learning} (HIOL), it is assumed that the human always labels the presented test cases correctly ~\cite{geethal2023human}\cite{grammar2fix}. The reason behind this assumption is that a person who knows the expected behaviour of the program under test (e.g. the developer or an end user) participates in debugging the program. However, in reality, the human can make mistakes, which adversely affects human-in-the-loop test oracle training. \learntofix~\cite{geethal2023human}\cite{bohme2020human} is the first human-in-the-loop test oracle training and program repair approach that can be applied to programs taking numeric inputs. Its journal extension~\cite{geethal2023human} suggests that incorrectly labelled test cases negatively affect \learntofix from multiple perspectives. This study addresses this issue by introducing a method called \afterofix to isolate mislabelled test cases introduced during the oracle learning process. According to the results in Section~\ref{sec:res}, \afterofix effectively identifies incorrectly labelled test cases introduced in \learntofix active oracle learning for many real-world bugs.

Several approaches have been proposed to identify incorrectly labelled data points in active learning~\cite{settles2009active}, as described in Section~\ref{sec:lit_review}. In common, these approaches require a large number of data points. This requirement is highly unlikely to be met in a human-in-the-loop environment. Also, noisy labelled detection approaches focusing on pool-based active learning approaches assume the existence of a set of unlabelled data points. In contrast, \afterofix does not require any additional or large test suites, making it suitable for a human-in-the-loop learning environment. 

In the software industry, \learntofix can be applied retroactively to generate an automated test oracle for a semantic bug, along with its fix~\cite{geethal2023human}. If a mislabelled test case is suspected to have been introduced during the \learntofix learning process, \afterofix can be executed to verify it. \afterofix can be executed multiple times, if necessary, to confirm whether mislabelled tests were introduced even in relabelling. The relabelling can be performed by the same individual involved in \learntofix. When the same test case is presented twice, the individual tends to examine it more closely than before. Moreover,  \learntofix can be converted into a \emph{pair programming strategy}~\cite{begel2008pair} through \afterofix by involving another individual or a group of individuals to answer the relabelling queries. If it is possible to find a \emph{reference implementation}~\cite{mechtaev2018semantic} of the program under test, it serves as an ideal source for the relabelling process of \afterofix.

The \learntofix journal extension~\cite{geethal2023human} reveals that the repairability and patch quality in automated program repair decrease due to the incorrectly labelled test cases introduced in the training process. Compared to \emph{Angelix}~\cite{mechtaev2016angelix}, the decrease in repairability in \emph{GenProg}~\cite{le2011genprog} is more significant. The \emph{fault localization}~\cite{wong2016survey} method used in GenProg considers a line executed only by failing tests to be highly suspicious, i.e., the fault can exist in that line. Thus, GenProg overlooks the faulty lines if the repair test suite contains incorrectly labelled failing tests (i.e., failing tests incorrectly labelled as passing). As \afterofix has a significant ability to identify incorrectly labelled failing tests (Section~\ref{sec:res_rq1}), the aforementioned issue can be fixed using \afterofix, improving both the repairability and patch quality in GenProg. Angelix can also be improved  similarly by \afterofix.

\afterofix can only be applied to programs taking numeric inputs. In future work, similar methods will be explored to detect incorrectly labelled test cases for human-in-the-loop oracle training methods for programs that take string inputs (E.g. \grammartofix~\cite{grammar2fix}). This work mainly focuses on identifying incorrectly labelled test cases introduced in HIOL. In future work, the methods for refining automatic test oracles compromised by mislabelled test cases will be explored. Also, the impact of correcting mislabelled test cases on the quality of automatic test oracles and automated program repair will be investigated. 

%% file: sections/conclusion.tex
\section{Conclusion}
This work introduces \afterofix to isolate \emph{incorrectly labelled test cases} introduced in \emph{human-in-the-loop} oracle learning for programs taking \emph{numeric inputs}. Given a compromised automatic test oracle and the training test suite used to train it, \afterofix systematically identifies such test cases based on their level of disagreement with others. Each test case should contain a fixed number of inputs and outputs for \afterofix to function properly. The experiments in this work demonstrate that \afterofix can accurately identify incorrectly labelled test cases introduced in \learntofix, the first and only human-in-the-loop oracle learning technique for programs taking numeric inputs. Moreover, \afterofix can effectively reduce the search space for identifying incorrectly labelled test cases. These capabilities of \afterofix mitigate the impact of mislabelled test cases on human-in-the-loop oracle learning and test-driven automated program repair.